\documentclass{article}
\usepackage[utf8]{inputenc}
\usepackage{amsmath}
\usepackage{amsfonts}
\usepackage{xcolor}
\usepackage{subcaption}
\usepackage{float}
\usepackage[top=1in, bottom=1in, left=1in, right=1in]{geometry}
\usepackage{amsthm}
\usepackage{mathtools}
\usepackage{graphicx}
\usepackage{indentfirst}
\usepackage{url}
\usepackage{algorithm}
\usepackage{algpseudocode}
\usepackage{bbm}
\usepackage{empheq}
\title{Toward Textual Transform Coding\footnote{Based on a lecture entitled ``Learning from Humans How to Improve Lossy Data Compression'',  given at the International Symposium on Information Theory in Espoo, Finland, summer of 2022.   }}
\date{\today}
\author{ 
Tsachy Weissman \\
  Department of Electrical Engineering \\
  Stanford University \\
  \texttt{tsachy@stanford.edu}
}

\usepackage{hyperref}
\usepackage[
backend=biber,
style=alphabetic,
sorting=none,
maxbibnames=99
]{biblatex}
\theoremstyle{definition}

\numberwithin{thm}{section}

\begin{document}

\maketitle

\begin{abstract}
   Inspired by recent work on compression with and for young humans,   the success of transform-based approaches to information processing, and  the rise of powerful language-based AI,  we propose  \emph{textual transform coding}. It shares some of its key properties with 
   traditional transform-based coding  underlying much of our current multimedia compression technologies. It can form the basis for compression at bit rates until recently considered uselessly low,  and for boosting human satisfaction from reconstructions at more traditional bit rates.   
   \end{abstract}


\section{Context and Motivation} 
\label{sec:context and motivation}

\subsection{State of our Compressors}
\label{subsec: The State of our Compressors}


Are  our  technologies for compression and streaming of audio, image and video approaching some kind of a     
rate-distortion-complexity Pareto front, or is there substantial room for improvement?  This question was brewing in my group and elsewhere throughout the last decade, in light of the growing appetite for capturing, storing and communicating such data, along with the emergence  of even more voluminous data types such as $3D$ point clouds and multi-sensor data from self-driving vehicles.    



\subsection{What would Shannon do?} 
\label{subsec:Shannon}
Following the publication of 
\cite{shannon}, Shannon directed his attention to the compressibility of English. Among other insights,  \cite{6773263} suggested that English text should be compressible to about $1.3$bits/character once our algorithmic predictive models catch up with those in Mrs.\ Mary E.\ Shannon's brain.  
Subsequent work \cite{1055912, 2019Entropy} has shown that number to be largely consistent across English speaking humans. 
It took our technologies over seven decades to  catch up and deliver \cite{NNCP, CMIX}.  

Inspired by this progression, 
we wanted to understand from humans what might be achievable 
in the context of multimedia data compression  once:
\begin{itemize}
    \item our algorithmic models catch up to those in our brains

    \item our codebooks/encoding/decoding make good use of humanity’s publicly available “side information” 

    \item we tailor the reconstructions to what humans actually care about 
\end{itemize}
In the summer of 2018 we were fortunate to host 3 high school students for an internship dedicated to an attempt at  addressing these questions. 

\subsection{Image Compression with Human Encoders, Decoders and Scorers}
\label{subsection: human image compression}
One intern (human encoder) was given a photo they hadn't previously seen and that was not available online. Their task was to describe it, in text, to another intern (human decoder) using the inbuilt Skype
text chat (turning off their outgoing audio/video).  
The decoder 
could share partial, in-progress reconstructions with the human encoder using
Skype’s screen share feature\footnote{The intermediate reconstructions are based on bits flowing from the encoder and thus  such ``feedback'' schemes are legitimate lossy compressors.}.
We compared the quality of the reconstructions under the human compression system to the quality under WebP using human scorers. Human compressors achieved higher quality than WebP confined to a similar bit rate on most  image types \cite{8712697, bhown2019improved, IEEEspectrum}. 

\subsection{SHTEM} 
\label{subsection: SHTEM}
Motivated by that project, in the summer of 2019 we launched the SHTEM summer internship program for high school and community college students \cite{SHTEM}. The H stands for both Humanities and the Human element, with an emphasis on multi-disciplinary projects combining humanities with STEM, and that use humans to assess, guide and evaluate what our technologies should strive to achieve. 

A couple of the SHTEM 2019 projects  continued the project from the preceding summer:  
One  was dedicated to gleaning insight into the potential for better facial image compression from the way in which a police sketch artist creates an image \cite{SHTEMfacial2019}. Another eliminated the human at the decoder from the setting of \cite{8712697} while reducing the bit rate by another order of magnitude without compromising human satisfaction. This was done  by tasking the human encoder with describing the image in Python code in a way that would allow a computer running it to create the reconstruction \cite{HAACwithpython}.

\subsection{Video}
\label{subsec: Video}
One of the SHTEM projects from the summer of 2020 was instrumental to  
what became StageCast \cite{stagecast}, a project born during the early days of the pandemic which explored and developed technology enabling real time performances with geographically distributed members of the cast and audience. 
The idea was to detect and stream the location of key points in the face and body of a human and have their digital puppet reconstructed by the decoder \cite{RoshanDCC2021}. The substantial reduction in the required bit rate enabled essentially real-time interactions that were not previously achievable with existing (full video streaming) technologies. That project also suggested the potential for a similarly low bit rate scheme streaming real video, replacing the digital puppet by a deep fake \cite{DEEPFAKE} of the original person at the reconstruction. This suggestion was evidently heeded by Nvidia in \cite{Nvidia}.  

With appetites whet, in \cite{TEXT2VID} we pursued a framework for video-conferencing compression at yet lower bit rate regimes. 
The gist was the realization that in a typical teleconference video, once the background and the person speaking have been learnt, inclusive of how that person sounds and  moves according to the content of their words, information theoretically the main part that remains is the content of their words. We devised a video streaming system exploiting existing technologies (such as for extracting text from audio and for synthesizing speech from text), which achieved user satisfaction levels similar to existing standards while requiring three orders of magnitude lower bit rates. 
Our SHTEMers implemented a version that runs on a standard web browser \cite{SHTEMDCC2023}. 

In \cite{Jenya} we used human input to teach a small deep net to anticipate regions of importance in a video, guiding the bit rate allocation to boost the performance of an existing video codec (x264). Naturally, the importance regions learned corresponded to objects that can be tagged by one or a couple of words. 

\subsection{Emergence of Text}
\label{subsec: Emergence of Text}
This work, performed largely by SHTEMers and their mentors, has revealed  
much potential for improvements over our existing technologies for multimedia data  compression, with  
human language emerging as key for assessing this potential and delivering on it.   
Meanwhile,   natural language processing and machine learning have been progressing dramatically, with the rise    
of powerful language models (such as ChatGPT, LaMDA, Bard, and Copilot), generators of multimedia data based on textual descriptions (such as DALL-E2), generators of text descriptions of multimedia data (such as in GPT4), etc.

\section{Quest for a New Transform}
\label{sec: Quest for a New Transform}
Our most widely used information processing technologies are transform based: transform, process-in-the-transform-domain, inverse-transform. 
Most widely used have been the linear tansfroms (FFT, wavelets, etc.), with non-linear transforms  recently playing an increasingly prominent role  \cite{nonlinear_transform_coding, attentionalluneed}.

\subsection{Effective Transforms}
\label{subsection: effective transform}
An effective transform in a given application has all or most of the following characteristics: 
\begin{itemize}
    \item coefficients in the transform domain correspond to elements (basis functions) that are  meaningful (biologically, physically, perceptually, or conceptually)
    
    \item sparsity of and simple relationships between transform coefficients (uncorrelated, independent, weakly dependent, etc.)
    

    \item smoothness of the forward and inverse transforms with respect to relevant metrics (i.e., two ``similar'' inputs remain ``similar'' in the transform domain, and vice versa)

\item low complexity (of both the forward and inverse transform)   
    \end{itemize}
An added  benefit of such transforms is that they induce wieldy yet  realistic  
data models.

\subsection{FFT as an Allegory}
The most celebrated example of an effective transform is that named after Fourier.   
A seasoned audio engineer, on a quick glance of  
 the Fourier transform of a segment of audio, can tell whether it's a musical piece and, if so, which instruments are playing what notes, the level of the background noise, etc.  
  Peaks in the transform domain correspond to dominant frequencies, in turn corresponding to ``harmonics'', i.e.\  signal components resonant with our auditory system. 
 
A harmonic in the Fourier context is meaningful to us: we know what it looks like on an oscilloscope, what it feels (sounds) like to our auditory system, and we have a (numerical) way of labeling it (frequency). What are our conceptual “harmonics”?  What ``thing'' lights up non-trivial subsets of our neurons?     

 The answer, one might argue, are our words, which  refer to and describe concepts or things that  resonate with us.  In the context of image data, our quest might benefit from the following analogy:  
 \begin{center}
\begin{tabular}{ c|c } 
 \textbf{Fourier Transform} & \textbf{New Transform}  \\ 
 \hline
 harmonics & objects for which we have words  \\ 
 \hline
 frequencies  & words  \\ 
 \hline
 amplitude and phase of harmonics  &	size,location, and orientation of objects  \\ 
 \hline
  precision of amplitude and phase  &	 precision of size and location  \\ 
 \hline
 frequency resolution &	number of words per object  \\ 
  \hline
component-wise fidelity & fidelity to story being told  \\ 
 \hline
FFT	& computationally efficient 
image captioning \\
\end{tabular}
\end{center}
Words are as central to the transform we seek as frequencies are to that of Fourier. 

\subsection{Qualitative Examples}
\begin{figure}
     \centering
     \begin{subfigure}{0.45\textwidth}
         \centering
         \includegraphics[width=\textwidth]{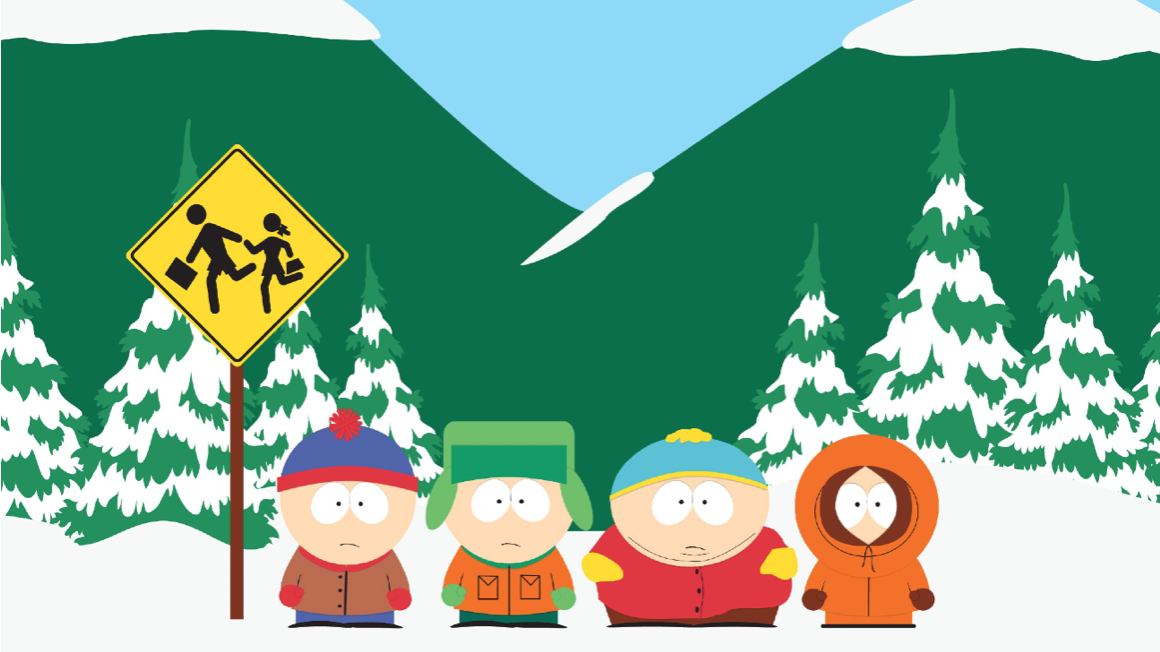}
         \begin{minipage}{.1cm}
            \vspace{0.6in}
         \end{minipage}
         \caption{in the original domain}
         \label{fig:original south park}
     \end{subfigure}
     \hfill
     \begin{subfigure}{0.45\textwidth}
         \centering
         \includegraphics[width=\textwidth]{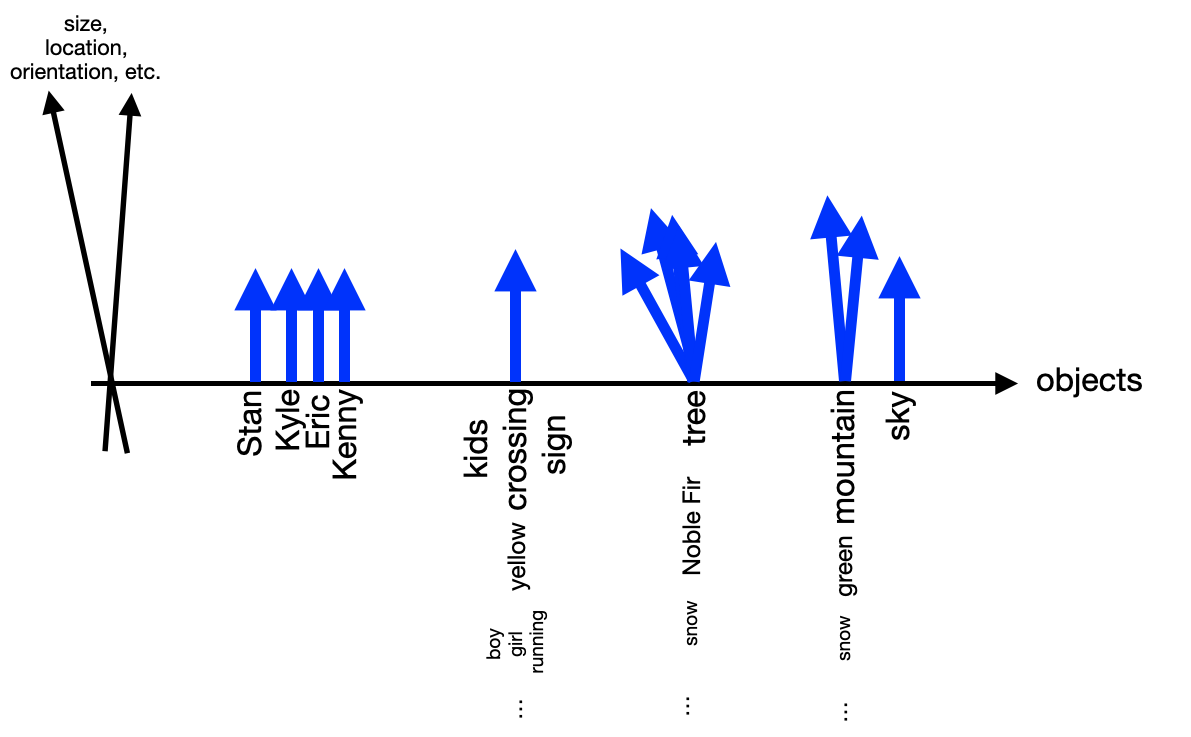}
         \caption{in a textual transform domain}
         \label{fig:south park image in textual transform domain}
     \end{subfigure}
        \caption{South Park image}
                \label{fig:south park image}
\end{figure}

It's instructive to contemplate what the transform we seek might look like in a couple of familiar edge cases. 
Let's begin with white noise. It remains white noise under Fourier or any of the other orthonormal transforms such as wavelets. But in the new transform domain it would be totally concentrated (a ``delta function'') on one element: white noise.  
\begin{figure}
     \centering
     \begin{subfigure}{0.45\textwidth}
         \centering
         \includegraphics[width=\textwidth]{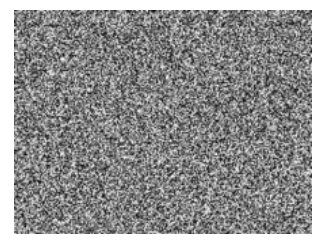}
         \begin{minipage}{.1cm}
            \vspace{0.6in}
         \end{minipage}
         \caption{in a traditional transform domain (remains white noise)}
         \label{fig:white noise in original domain}
     \end{subfigure}
     \hfill
     \begin{subfigure}{0.45\textwidth}
         \centering
         \includegraphics[width=\textwidth]{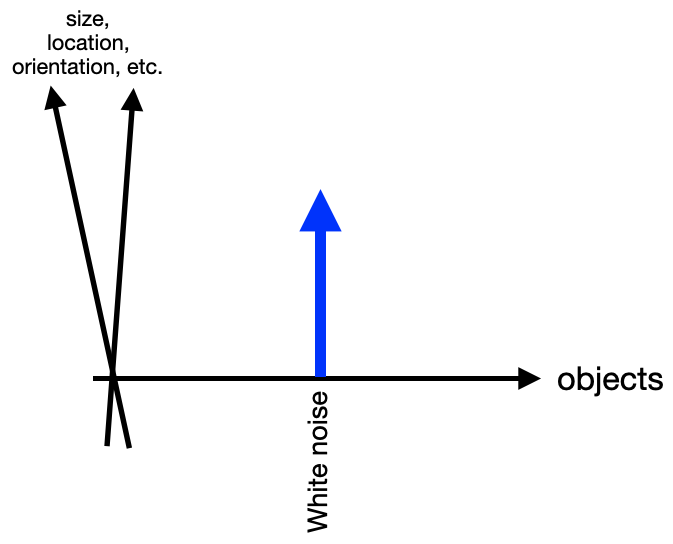}
         \caption{in a textual transform domain}
         \label{fig: white noise in textual transform domain}
     \end{subfigure}
        \caption{white noise 
        }
        
        \label{fig:white noise}
\end{figure}
Next let's consider the South Park  image in Figure \ref{fig:south park image} and a cartoon of what a version of it might look like in such a transform domain. 
The point  is that there are image types    -- where the story being told is more important than the pixel-wise fidelity -- that would have a sparse approximate representation under this transform. These might include holiday greetings cards,  wedding photos, cartoons,  
computer-generated images, 
propaganda posters and ads.

\subsection{A Textual Transform}
\label{subsec: a textual transform}
Moving from the qualitative to the concrete, a possible version  of a (lossy)   
representation of an image in a transform domain of the type alluded to in Figure \ref{fig:white noise} and Figure \ref{fig:south park image} would comprise the following elements: 
\begin{itemize}
    \item   description of how many objects (up to prescribed number $L$ for cognitive load) 
    \item list of what they are (one word each) 
    \item their sizes, locations, orientations (at prescribed physical resolution $R$)
    \item  words of description for each and how they relate to each other (at prescribed `textual resolution' $W$)
\end{itemize}
Different choices of $(L, R, W)$  correspond to different bit rates and description quality. 
Each of these elements can be described exclusively in text, yielding what boils down to a verbal description of the image, confined to a given budget of words per object, sentences, overall length, etc. 

Such a transform seems to check the boxes of Subsection \ref{subsection: effective transform}, inclusive of ``smoothness'' with respect to textual similarity metrics (cf.\   \cite{textsimilaritysurvey} and references therein) and 
``low complexity''. 
Technology for implementing versions of such transforms (image captioning) and their inverses (generating images from text) is rapidly evolving   \cite{imagecaptioningpaper, Galatolo_2021}.

Such a transform is particularly useful for lossy compression.  It 
 comprises the compressed representation, 
as in the setting of Subsection \ref{subsection: human image compression}, sans the human in the loop. 
 Beyond checking the boxes of a good transform, this one comes with another level of meaningfulness of the compressed representation, which is human readable and queryable. 
Here encoding and decoding are each of standalone value: the encoding being a human readable textual summary of the data, and the decoder being a generator of a point in the original data domain based on its textual description. 

\begin{figure}
     \centering
     \begin{subfigure}{0.30\textwidth}
         \centering
         \includegraphics[width=\textwidth]{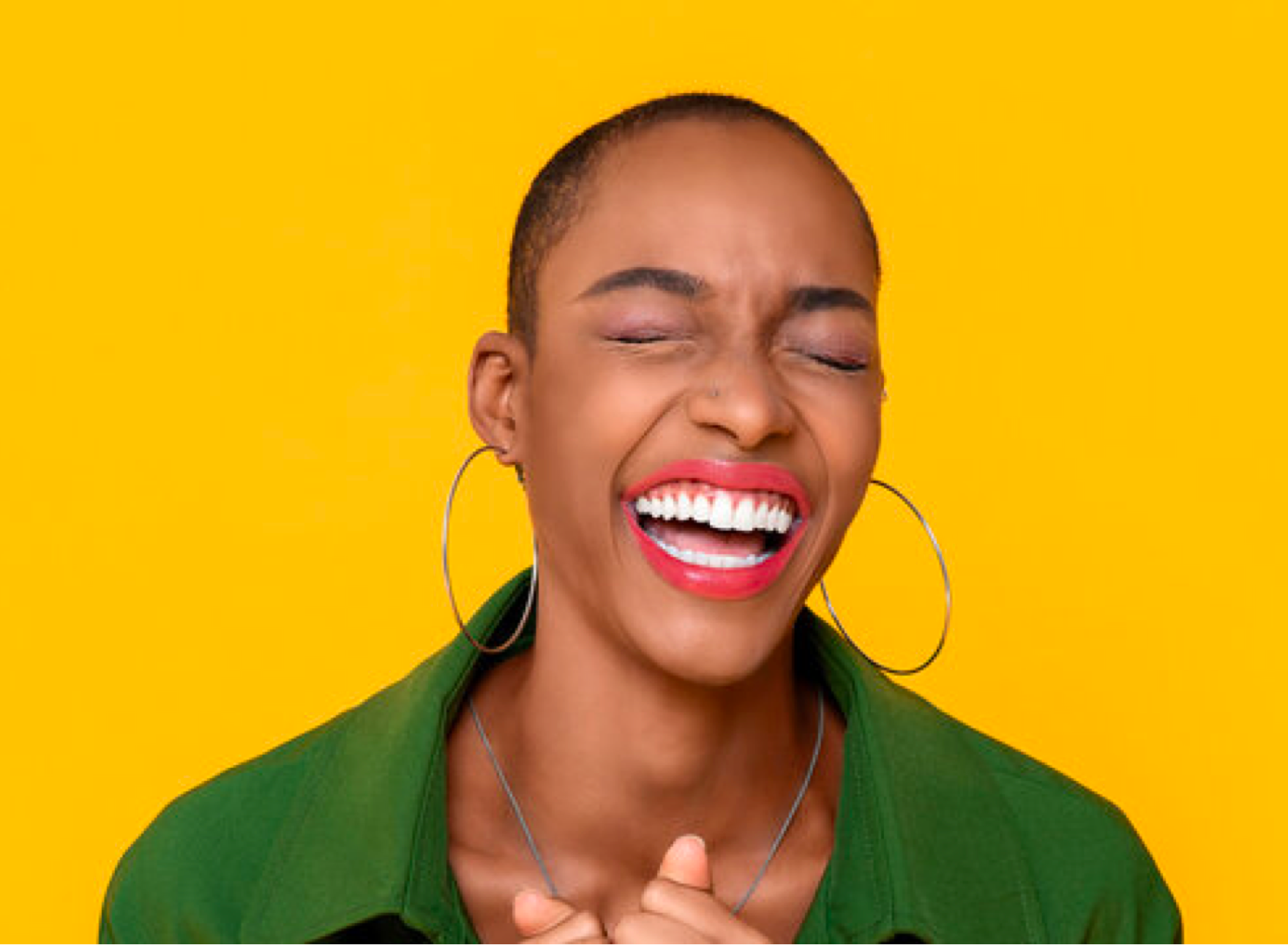}
         \caption{source data}
         \label{fig:source data}
     \end{subfigure}
     \hfill
     \begin{subfigure}{0.30\textwidth}
         \centering
         \includegraphics[width=\textwidth]{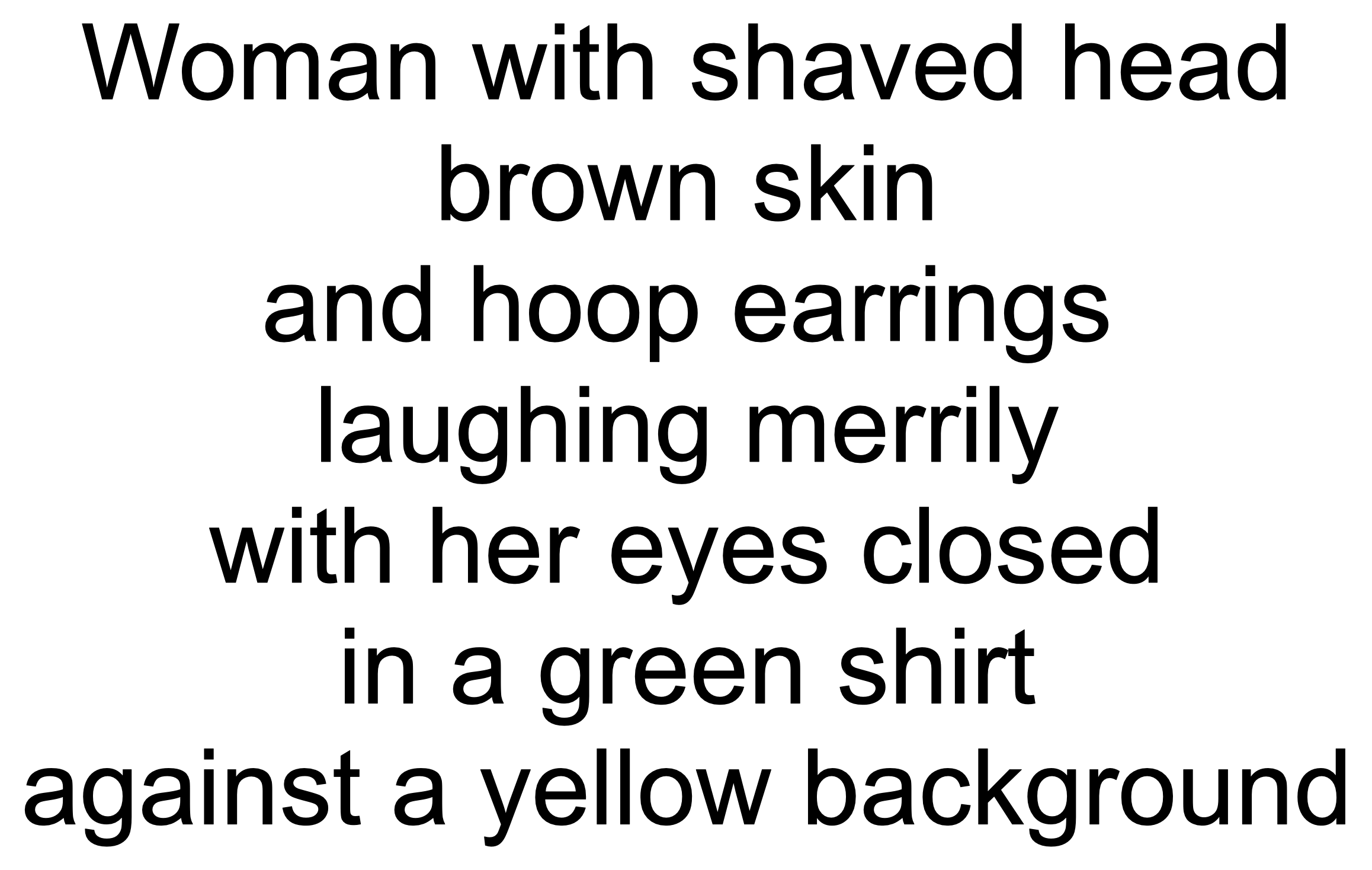}
         \caption{textual representation}
         \label{fig:textual representation}
     \end{subfigure}
     \hfill
     \begin{subfigure}{0.30\textwidth}
         \centering
         \includegraphics[width=\textwidth]{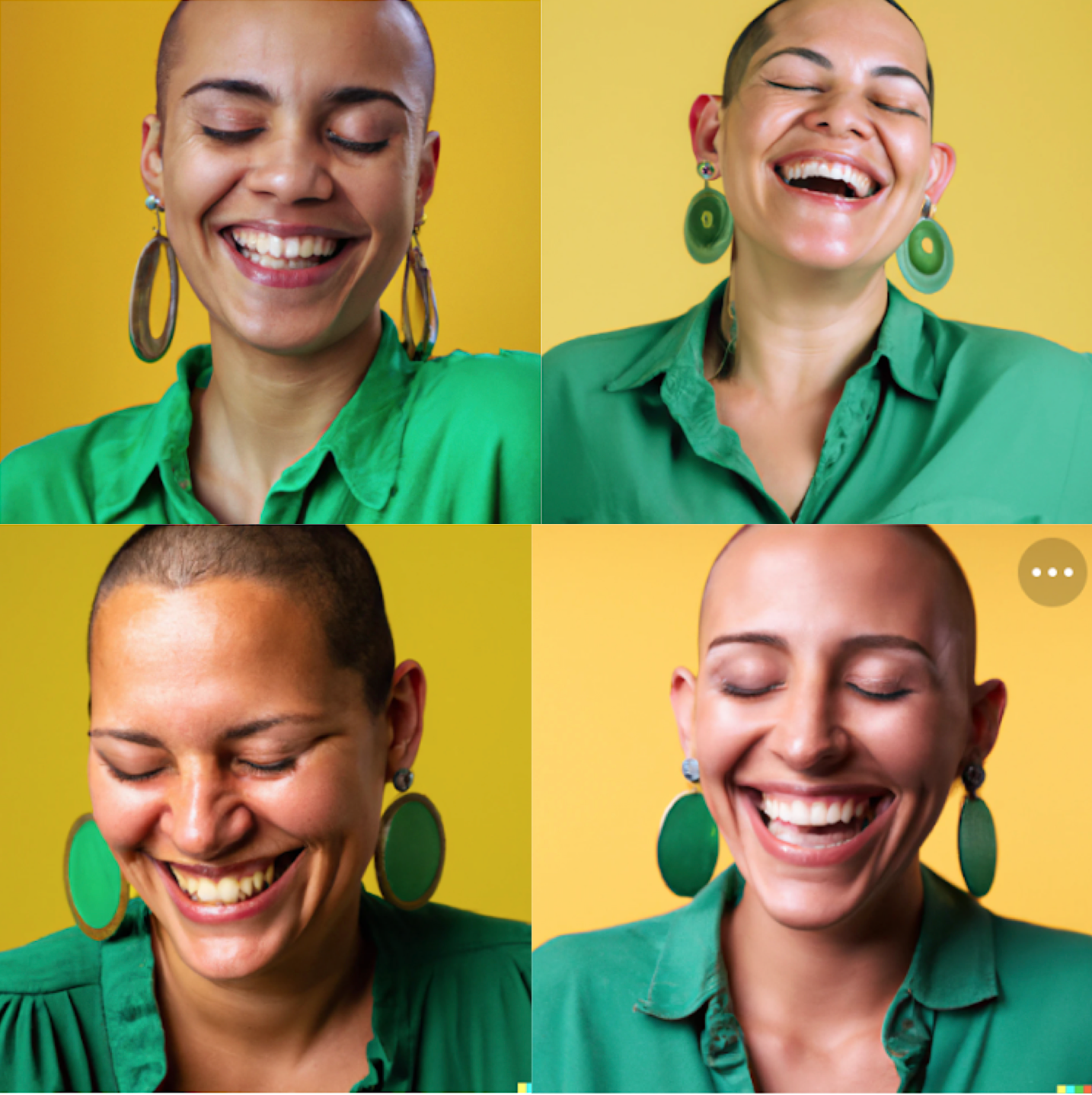}
         \caption{DALL-E2 reconstructions}
         \label{fig: DALL-E2 reconstructions}
     \end{subfigure}
        \caption{Extreme compression via textual transform coding}
        \label{fig: Extreme compression via textual transform coding}
\end{figure}

Figure \ref{fig: Extreme compression via textual transform coding} depicts an extreme version of this idea, implementable with existing technologies, highlighting that the compression rate, in the traditional sense of bits per symbol (pixel), can be made ludicrously low 
with reconstructions that might be satisfactory in some applications. Perhaps more meaningful is
that the overall number of bits required for the compressed representation, along with the complexity of the implementation,  scale with the amount of nuance in the story being told rather than
the physical (pixel) resolution of the image.

\subsection{Classical vs. Textual regimes}
\begin{figure}
     \centering
     \begin{subfigure}{0.30\textwidth}
         \centering
         \includegraphics[width=\textwidth]{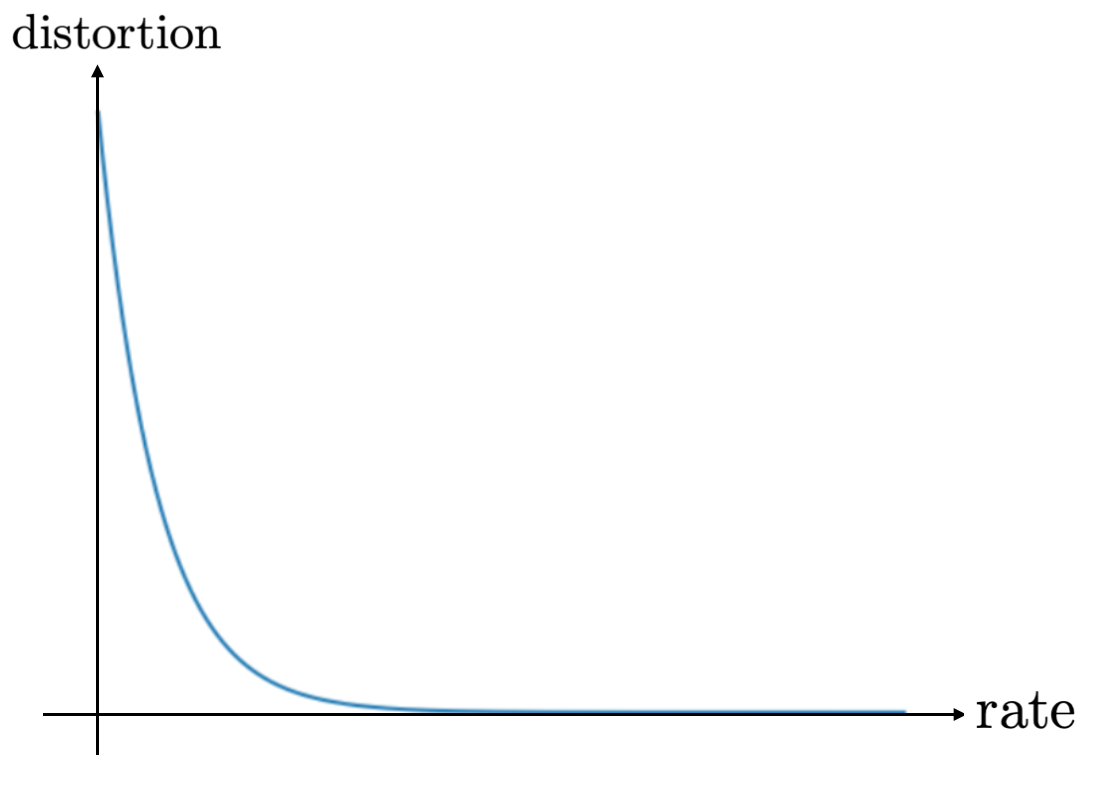}
         \caption{Distortion-Rate function}
         \label{fig:Distortion-Rate function}
     \end{subfigure}
     \hfill
     \begin{subfigure}{0.30\textwidth}
         \centering
         \includegraphics[width=\textwidth]{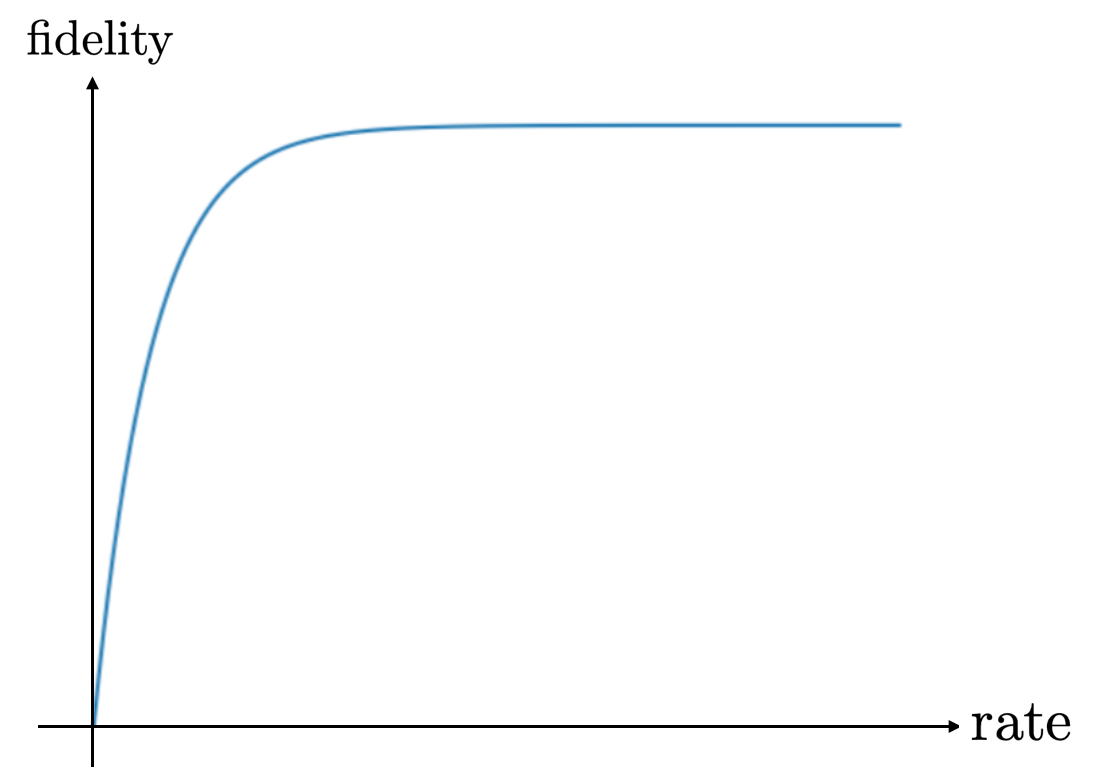}
         \caption{Fidelity-Rate function}
         \label{fig:Fidelity-Rate function}
     \end{subfigure}
     \hfill
     \begin{subfigure}{0.30\textwidth}
         \centering
         \includegraphics[width=\textwidth]{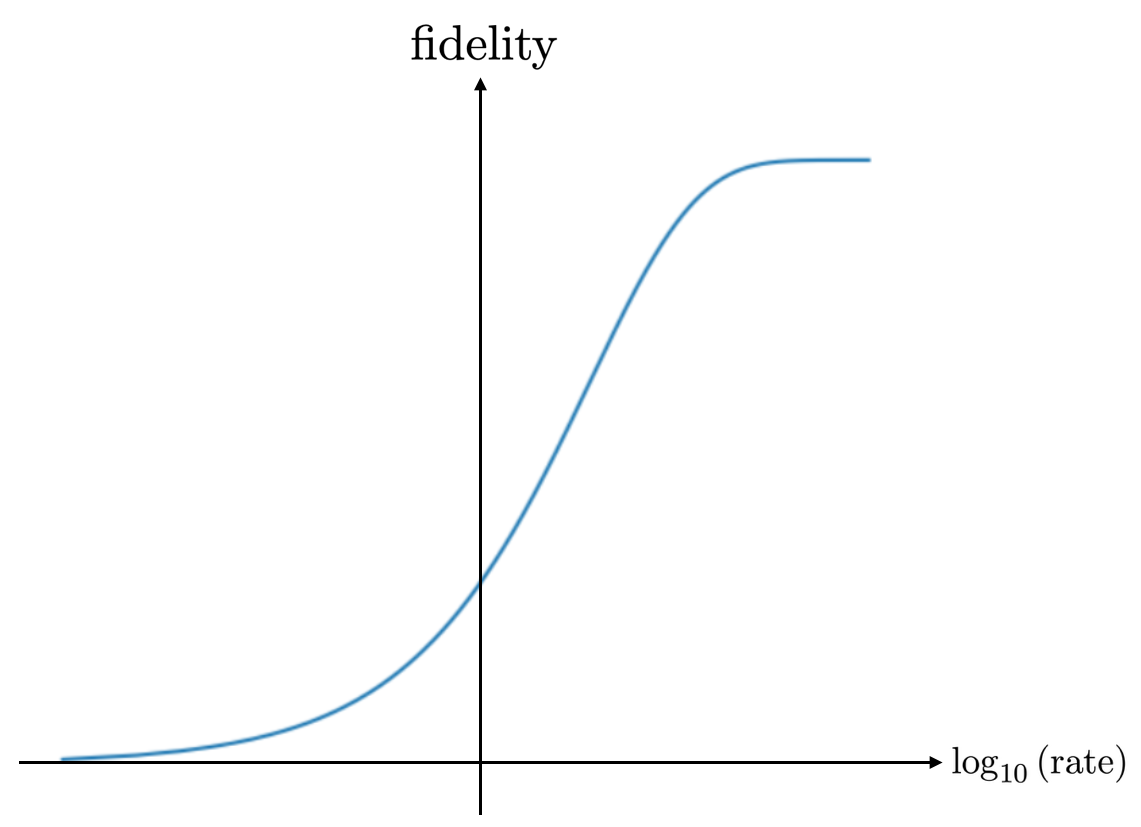}
         \caption{Fidelity vs. $\log_{10}$(Rate) }
         \label{fig: Fidelity vs log Rate}
     \end{subfigure}
        \caption{Distortion-Rate and Fidelity-Rate functions under classical symbol- or pixel-wise criteria}
        \label{fig: Distortion-Rate and Fidelity-Rate functions under symbol- or pixel-wise criteria}
\end{figure}

Shannon's Rate-Distortion theory has focused on symbol-by-symbol distortion criteria. 
Figure \ref{fig:Distortion-Rate function} depicts the Distortion-Rate function of a memoryless Gaussian source under squared error distortion as representative of what will qualitatively look quite similar for any non-degenerate source and symbol-by-symbol distortion criterion. Rate is measured in bits per source symbol, commensurate with a regime of an overall number of bits linear in the number of source symbols being represented. Sufficiently large rate achieves arbitrarily small distortion (in fact zero for discrete sources when reaching or exceeding the entropy rate). Equivalently, one might define fidelity as the negation of distortion, with zero distortion corresponding to maximal fidelity. Figure \ref{fig:Fidelity-Rate function} depicts the  
Fidelity-Rate function that would result for the Distortion-Rate function of Figure \ref{fig:Distortion-Rate function}, with maximal fidelity set in this example to the variance of the source. Yet another equivalent representation is depicted in Figure \ref{fig: Fidelity vs log Rate}, with a logarithmic scale for the rate. It elucidates 
the fact that, in this framework of symbol-by-symbol distortion, extremely small information rates can at best only negligibly boost the fidelity from its minimal value.

\begin{figure}
     \centering
   
         \includegraphics[width=\textwidth]{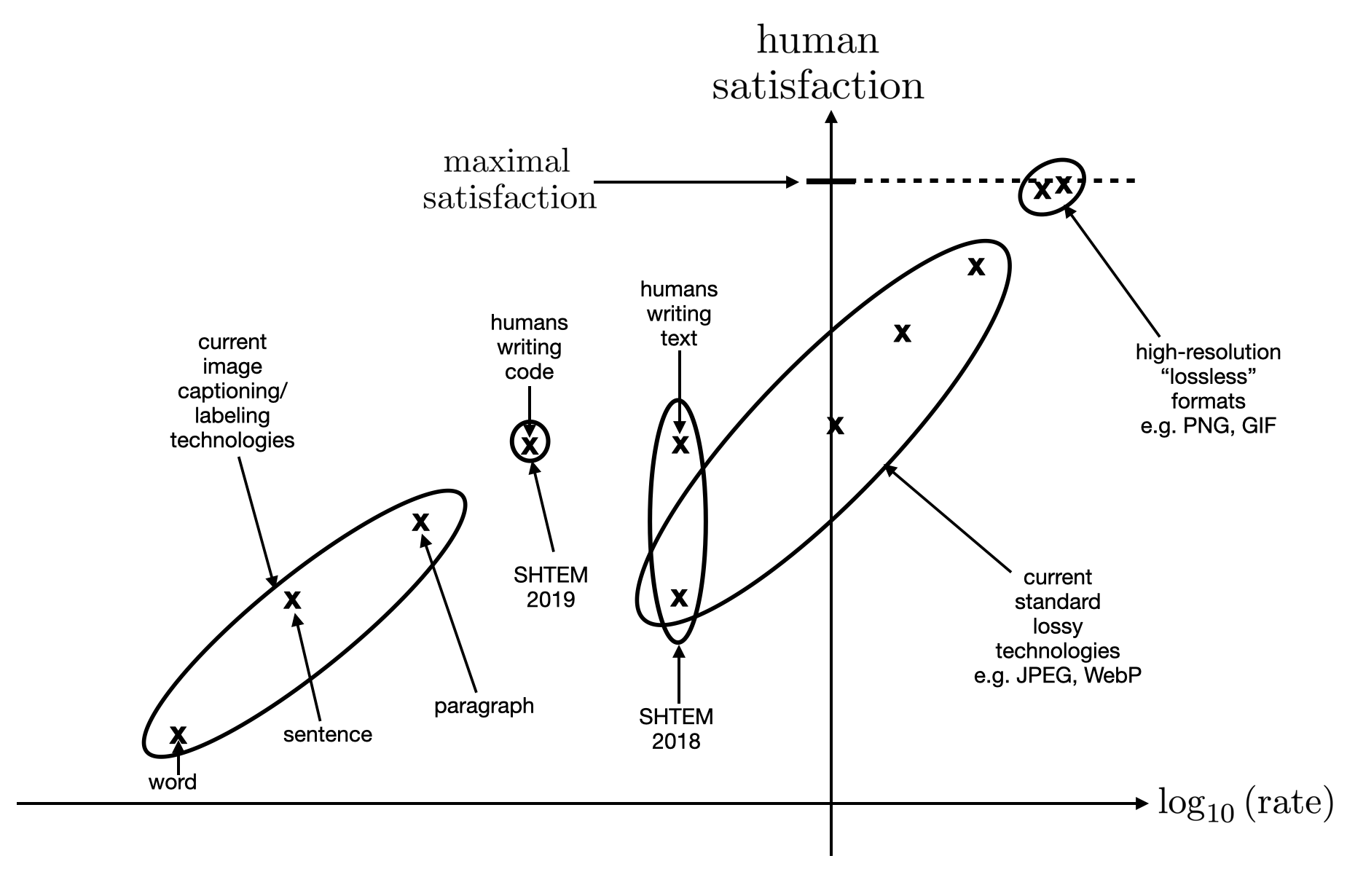}
                 \caption{Achievable human satisfaction levels}
        \label{fig: qualitative human satisfaction}
\end{figure}

This behavior is quite different from  that we have been observing of human satisfaction from text-based reconstructions in multimedia data compression, as depicted in Figure \ref{fig: qualitative human satisfaction}.  
Our initial experiment discussed in Subsection \ref{subsection: human image compression} has shown that extremely low bit rates -- well on the left in a plot with logarithmic rate on its $x$-axis -- suffice for substantially boosting human satisfaction levels beyond the minimal value. Subsequent work mentioned in Subsection \ref{subsection: SHTEM}
showed similar reconstruction quality achievable while reducing the rate by another order of magnitude. Our current experiments (not yet public) are suggesting that non-trivially positive human satisfaction levels are achievable with further orders of magnitude reductions in rate via textual transform coding at regimes of paragraphs, handful of sentences (as in Figure \ref{fig: Extreme compression via textual transform coding}), handful of words, etc.

\section{Next steps} 
\label{sec: Current and future work}
\subsection{A Fidelity Measure Aligned with Human Satisfaction} 
\label{subsec: A loss function aligned with human satisfaction}
We conjecture 
that human satisfaction, given a source and its possible reconstruction, in the context of multimedia data, is mainly determined by two factors. The first is the degree of fidelity to the story being told, as would be captured by  similarity in the textual domain. The second is fidelity in the traditional  sense. In other words, human satisfaction can be well approximated/predicted by a function of the form 
\begin{equation}
    s(x , \hat{x} )= g \left( T_f  , P_f \right) , 
\end{equation}
where $x$ is the source, $\hat{x}$ its  reconstruction, $T_f(x , \hat{x} )$ is similarity measured between the respective textual transforms ($T_f$ standing for ``textual fidelity''), and $P_f(x , \hat{x} )$ a traditional type of similarity measure such as PSNR or SSIM ($P_f$ standing for ``pixel-wise fidelity'').  
Figure \ref{fig: qualitative human satisfaction} suggests that high satisfaction can be achieved primarily via   high traditional fidelity and would be largely determined by it in that regime. In the other extreme, when traditional fidelity is low, the level of satisfaction would be largely determined by the level of fidelity in the textual domain. In other words,  
assuming without loss of generality that $T_f$ and $P_f$ are normalized to reside in $[0,1]$,  
$g \left( \alpha  , \beta \right)$ would be largely dominated/determined by $\alpha$ for $\beta$ low and $\beta$ when $\beta$ is high.   We plan to dedicate a couple of SHTEM 2023 projects to gauging and quantifying human satisfaction, and to experimentally assessing the validity of our conjecture along with the form of the function $g$.

\subsection{Characterizing and Approaching $S(R)$} 
\label{subsec: satisfaction rate theory charcterized}
As alluded, the textual transform would be a useful framework for modeling the data. Unlike  symbol-by-symbol approaches which tend to yield models either too simplistic for the actual data or too complicated to be useful, one could start in the textual domain, which is amenable to fairly simple and realistic modeling, followed by a generative model going from the text to the original domain. With such an approach, and the satisfaction function discussed in the preceding subsection identified, it would be realistic to characterize the best achievable trade-offs between satisfaction and (log of the) rate by applying the well-developed tools from the Shannon theoretic arsenal in both the textual and generative domains.

\begin{figure}
     \centering
   
         \includegraphics[width=\textwidth]{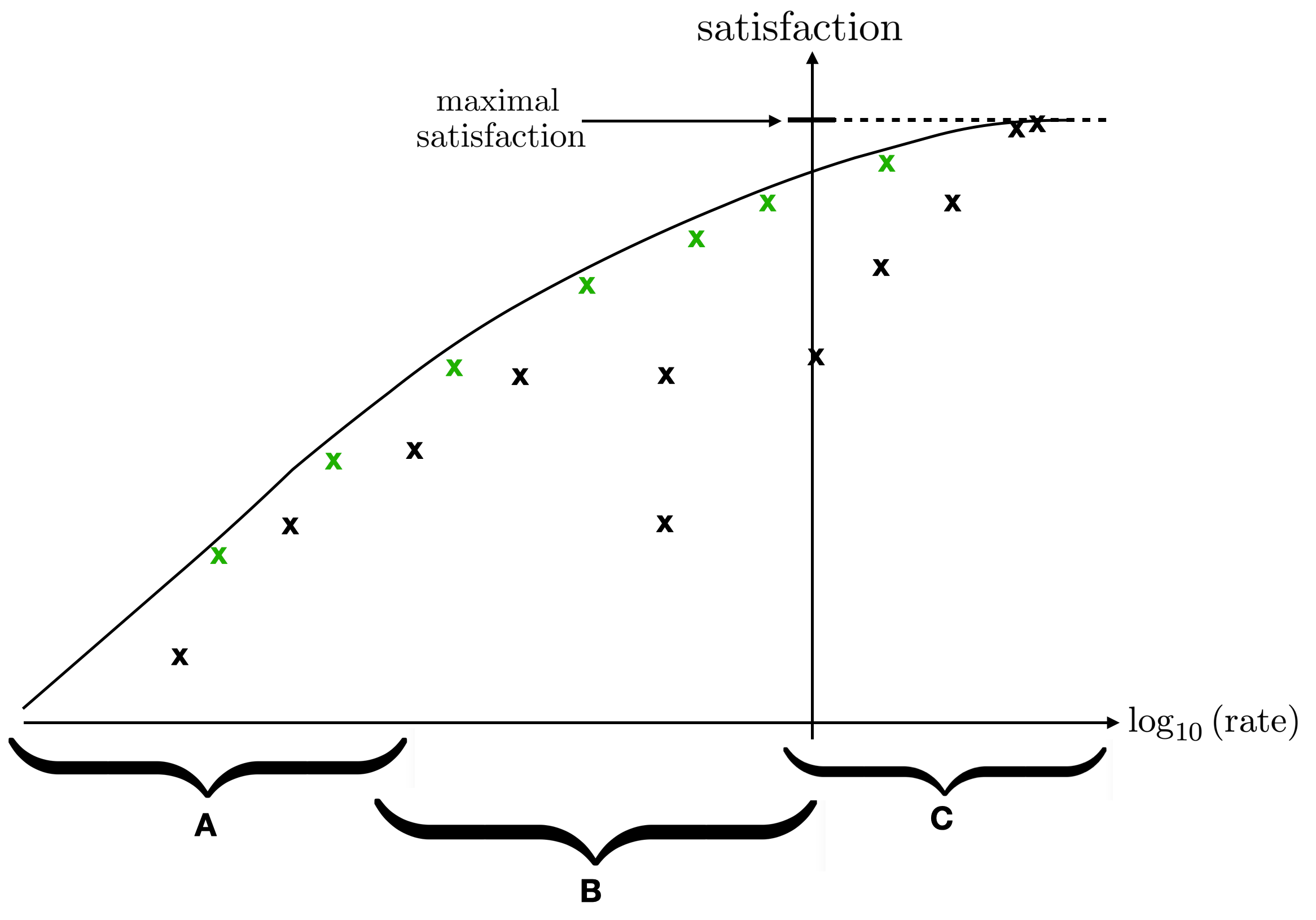}
                 \caption{$S(R)$ represented by the solid curve, black points are from Figure \ref{fig: qualitative human satisfaction}, green points represent schemes to be developed. A is the ultra-low rate region where textual transform coding is key, C is the region where most compression technologies currently operate, B is a largely unexplored intermediate region necessitating new approaches for bridging the textual with the traditional regimes.}
        \label{fig: satisfaction rate function and approaching it}
\end{figure}

On the constructive side, characterizing $S(R)$ would accompany and guide the development of new practical schemes. 
Progress on textual technologies of the type mentioned in Subsection \ref{subsec: Emergence of Text} will directly translate to improved schemes for the ultra low rate region. No less interesting would be the largely unexplored region bridging the ultra-low and the traditional symbol-by-symbol (pixel-wise) fidelity region where most current technologies reside. This region will likely benefit from new approaches for effectively leveraging and combining information from the textual domain with low resolution versions from the more traditional ones, as suggested in 
 Figure \ref{fig: satisfaction rate function and approaching it}.

\subsection{Beyond Compression}
\label{subsec: Beyond compression} 

\subsubsection{Textual Transform and Privacy}
\label{subsubsec: TT and privacy} 
The ultra low-rate regime enabled by textual transform coding   
could be beneficial not only for the dramatic space/bandwidth savings but also privacy. Representing the source data via text that can be compressed into a handful of $k$ bits  
translates trivially to a privacy guarantee that the mutual information between the image and its representation is upper bounded by $k$. A variant is to let those $k$ bits represent answers
to a set of queries about the source data that would be agreed upon as non-private, naturally extractable from the textual domain. Recent work in this direction has been reported on in \cite{guo2022protecting}.  

\subsubsection{Denoising}
\label{subsubsec: TT and denoising} 
One could envision applications and data types where denoising can be performed particularly dramatically and effectively in the textual domain. 
E.g., simply removing the word `noisy' from the textual description or replacing it with the word `crisp' could result in a
substantial quality boost, as can be measured and optimized for under the satisfaction function of  Subsection \ref{subsec: A loss function aligned with human satisfaction}.


\section{Conclusion}\label{sec:conclusion}
The textual transform  is a conceptually useful and increasingly practical framework for multimedia information processing based on a code optimized over many years of human evolution. 
It yields human readable and searchable representations,
with bounded implementation complexity that does not grow with traditional physical 
characteristics such as pixel resolution. 
It may be key to enabling compression and streaming at unprecedentedly low bit rates, and to characterizing 
fundamental trade-offs between bit rate and human satisfaction.    


Our focus here was images for illustrative purposes, 
 but the ideas are similarly applicable to other traditional and new multimedia data types. 
Our work joins  other recent activity attempting to incorporate notions of perception and semantics into traditional compression and communication paradigms, cf.\     \cite{blau2019rethinking, zhang2021universal, guo2022itw, guo2022semantic, Gunduz:2023:10.1109/JSAC.2022.3223408, agustsson2023multirealism} and references therein. Further progress in this area will likely come from multidisciplinary collaborations between information scientists, engineers, neuroscientists and psychologists.





\section{Acknowledgement}
I am grateful to the many who have been making our SHTEM program a growing success. In particular, to  Cindy Nguyen, Sylvia Chin and Suzanne Sims for their tireless selfless work running everything since the summer of 2019; to Dr.\ Shubham Chandak, Dr.\ Kedar Tatwawadi, Dr.\ Pulkit Tandon, 
Dr.\ Irena Hwang, Dr.\ Yanjun Han, Devon Baur, Noah Huffman, Dr.\ Dmitri Pavlichin,  
Jay Mardia and many other Stanford faculty and students for their exquisite mentorship of the projects; and to the SHTEMers for their inspiring work. Thanks to Lara Arikan for the example in Figure \ref{fig: Extreme compression via textual transform coding} and stimulating discussions.

\printbibliography


\end{document}